\documentclass[12pt,a4paper]{JHEP3}

\usepackage{graphicx}
\usepackage{amsmath}
\usepackage{dsfont}
\usepackage{slashed}
\usepackage{braket}
\usepackage{epsfig}
\usepackage{subfigure}
\usepackage[percent]{overpic}

\def\p{\partial}
\def\Lag{\mathcal{L}}
\def\Ham{\mathcal{H}}

\title{Corrections to Nambu-Goto energy levels from the effective string action}

\preprint{WIS/12/10-AUG-DPPA}

\author{Ofer~Aharony and Nizan~Klinghoffer\\\\
Department of Particle Physics and Astrophysics, Weizmann Institute of Science,
Rehovot 76100, Israel\\ \\
{\tt E-mails : Ofer.Aharony@weizmann.ac.il, Nizan.Klinghoffer@weizmann.ac.il} }

\abstract{
The effective action on long strings, such as confining strings in pure Yang-Mills theories,
is well-approximated by the Nambu-Goto action, but this action cannot be exact. The leading
possible corrections to this action (in a long string expansion in the static gauge), allowed by Lorentz invariance, were recently identified, both for closed strings and for open strings.
In this paper we compute explicitly in a Hamiltonian formalism the leading corrections to
the lowest-lying Nambu-Goto energy levels in both cases, and verify that they are consistent
with the previously computed effective string partition functions. For open strings of length $R$ the leading
correction is of order $1/R^4$, for excited closed strings of length $R$ in $D>3$ space-time dimensions
it is of order $1/R^5$, while for the ground state of the closed string in any dimension it is
of order $1/R^7$. We attempt to match our closed string corrections to lattice results, but
the latter are still mostly outside the range of convergence of the $1/R$ expansion that
we use.
}

\begin{document}

\bibliographystyle{utcaps}
\section{Introduction}

Many field theories in $D \geq 3$ space-time dimensions contain stable string-like excitations.
These include confining strings in Yang-Mills theories with no fundamental flavors,
Abrikosov-Nielsen-Olesen strings in the Abelian Higgs model, domain walls in $2+1$ dimensional
field theories, and so on. These strings have massless bosons on their worldsheet which are
Nambu-Goldstone bosons of the translation symmetries broken by the string, and in the absence
of other symmetries these are expected to be the only massless fields on the worldsheet. When
the field theory has a mass gap, one can thus write down a low-energy effective action for the
field theory in the presence of a long string, that includes only the massless bosons on the
string worldsheet; this effective action should be valid until the energy scale of any
additional degrees of freedom (on the worldsheet or in the ``bulk'' of space-time).

The effective action should be invariant under diffeomorphisms (reparameterizations of the
string worldsheet) and under Lorentz transformations (some of which are spontaneously broken
by the long string).
The simplest such action is the Nambu-Goto (NG) action, which in Euclidean space is just the string tension times the area of the worldsheet the string sweeps. Lattice simulations of $SU\left(N \right)$
Yang-Mills theories in $2+1$ and $3+1$ dimensions show that the flux-tube excitations have an energy spectrum close to that of NG, which suggests that the effective action for the string
can be written as the NG action with small correction terms.

In this paper we consider {\it long strings} obeying $R \gg 1 / \sqrt{T}$, where $R$ is the string length and $T$ is its tension. For an open string, we force it to be long by taking it to end on
two objects separated by a distance $R$ (for a confining string these can be two external quarks), while for a closed string we assume that one of the space-time coordinates is a circle of circumference $R$,
and we wrap the string around this coordinate once. We consider small excitations
around straight strings, and without loss of generality we assume that the string stretches
mostly along the $X^1$ direction. It is then convenient \cite{LW,AK} to use the static gauge
for the worldsheet coordinates, $\sigma^0 = X^0$, $\sigma^1 = X^1$, in which the degrees of
freedom of the effective action are simply the transverse positions of the string, $X^i(\sigma^0, \sigma^1)$ for $i=2,\cdots,D-1$.
The low-energy effective action is a derivative expansion on the worldsheet. For closed strings, the lowest terms
in this expansion which are allowed by the manifest $SO(1,1)\times SO(D-2)$ symmetry are:

\begin{equation}
S_E = S_{Free} + S_2 +S_3 + S_4,
\end{equation}
with
\begin{eqnarray}
\label{action}
 S_{Free} & \equiv & -\int d^2\sigma  \frac{1}{2} \p_\alpha X_i \p^\alpha X^i, \nonumber \\
S_2 & \equiv & -\frac{c_2}{4} \int d^2\sigma  \left( \p_\alpha X_i \p^\alpha X^i \right)^2,   \\
S_3 & \equiv & -\frac{c_3}{4}  \int d^2\sigma  \p_\alpha X_i \p_\beta X^i \p^\alpha X_j \p^\beta X^j, \nonumber \\
S_4 & \equiv & - c_4 \int d^2\sigma   \p_\alpha \p_\beta X_i \p^\alpha \p^\beta X^i \p_\gamma X_j \p^\gamma X^j. \nonumber
\end{eqnarray}
We use a Minkowski metric on the worldsheet with signature $(-,+)$, $\alpha ,\, \beta ,\, \gamma$ go over the worldsheet indices,
and $i ,\, j$ are space-time indices running over $\{ 2, \cdots, D-1 \}$. We didn't include in the action terms which are proportional to the
equations of motion as they can be removed by field redefinitions and thus they do not contribute to the energies; the minus signs in (\ref{action}) are
for consistency with our references.
Note that we can regard the derivative expansion as an expansion in a dimensionless parameter $ 1/ T R^2 \ll 1$. In (\ref{action}) we included only a single six-derivative term; there are also
other terms at the same order in the derivative expansion, but this term is the most interesting
one for reasons we will now explain.

While naively the coefficients $c_i$ should be generic, it was shown in \cite{LW,AK,AKS} that some of them are {\it constrained} by the Lorentz-invariance
of the field theory. This can be checked, for instance, by using open-closed-channel
duality and assuming Lorentz-invariant propagation in the closed channel. It was argued in \cite{LW,AK} that this requires $2c_2=-c_3$ and $c_3=-1/T$,
as one finds in the Nambu-Goto action, and that
for $D>3$ at six-derivative order the $c_4$ term in (\ref{action}) is the only allowed deviation from the Nambu-Goto action. For $D=3$ the leading allowed deviations from the Nambu-Goto
action arise at eight-derivative order \cite{AK}.

A similar analysis for open strings was recently performed in \cite{AF}; it was already shown in
\cite{LW} that no two-derivative boundary terms are allowed, and \cite{AF} showed that (assuming
Dirichlet boundary conditions $X^i=0$ for the transverse coordinates) there is a single allowed
boundary term at four-derivative order, of the form (assuming the boundaries are at constant
$X^1$)
\begin{equation}
 S'_2=b_2 \int_{boundaries} d\sigma^0 \, \p_{0}\p_{1}X^i \p_{0}\p_{1}X^i,
\end{equation}
for an arbitrary coefficient $b_2$.

In previous work on this subject the general form of the action was analyzed, but
corrections to specific energy levels, which are measured (say) in lattice simulations
of pure Yang-Mills theories, were not presented. More precisely, the partition functions
of the effective string theory on the annulus and on the torus were computed, and this includes (in the
derivative expansion) some information about the energy levels, but it only include the
sum of the energies of all states that are degenerate in the free theory $S_{Free}$, and
not corrections to individual energy levels.
In this work we calculate directly the leading corrections (compared to Nambu-Goto) to the low-lying energy levels
of closed and open strings, coming from the $c_4$ and $b_2$ terms above.
We start in section 2 by discussing the open string case, where we first work with the action $S_{Free}+S_2+S_3$ with {\it arbitrary} coefficients in-order to present our methods and compare
the results with the computations of \cite{LW}, which also computed the energy levels of this
action by a different method. We then discuss the leading corrections to Nambu-Goto which are
allowed, and verify their consistency by comparing with the known partition function on the annulus.
 In section 3 we describe the closed string case, computing the corrections coming from the
 $c_4$ term and verifying their consistency with the annulus and torus partition functions. We end in section 4 with a discussion of the extent to which our results can be compared to recent lattice results for
 $3+1$ dimensional gauge theories \cite{Barak,Teper}, showing some deviations from the Nambu-Goto energy levels.

\section{Open string energy levels}

We will use a Hamiltonian approach to compute the corrections to the energy levels, using a specific prescription for normal ordering and regularization of diverging sums. To verify
the consistency of our prescription, we begin in section \ref{sec:OpenC2C3} by taking arbitrary coefficients for the first two interaction terms in the effective action,
$S_{2,3}$~; we then successfully compare our results to the ones computed in \cite{LW}. In sections \ref{sec:OpenC4} and
\ref{sec:OpenB2} we compute the leading corrections to NG energy levels coming from the bulk ($c_4$) and boundary ($b_2$) terms, respectively, that are allowed to appear in the effective
action.

\subsection{The $c_{2,3}$ terms}
\label{sec:OpenC2C3}

We consider a strip-like string worldsheet $\sigma_0 \subset [-\infty, \infty]$, $\sigma_1 \subset [0,R]$, with
Dirichlet boundary conditions $X^i=0$ for the transverse coordinates at its ends, appropriate for a quark-anti-quark flux tube connecting external quarks. Note that we normalize the transverse scalars $X^i$ to be dimensionless, so the actual space-time distances are $X^i/\sqrt{T}$.
The general bulk
action before imposing the space-time Lorentz-invariance constraints, up to four-derivative order
(or $O\left( 1 / R^3 \right)$ corrections to energy levels), is~:
\begin{equation}
 \Lag=-\frac{1}{2}\p_{\alpha}X\cdot \p^{\alpha}X-\frac{c_{2}}{4}\left(\p_{\alpha}X \cdot \p^\alpha X\right)^{2}-
\frac{c_{3}}{4}\left(\p_{\alpha}X \cdot \p_{\beta}X\right)^{2},
\end{equation}
where the dot product indicates a summation over the transverse directions.
The conjugate momentum to $X^i$ is
\begin{equation}
\label{ConjMomentum}
\Pi^{i}=\frac{\delta \Lag}{\delta \left(\p_0 X _i\right) }=
\p_{0}X^{i}+c_{2}\left(\p_{\beta}X\right)^{2}\p_{0}X^{i}+c_{3}\p_{\beta}X^{i}\left(\p_{0}X^{j}\p_{\beta}X_{j}\right),
\end{equation}
and the Hamiltonian $\Ham=\int d\sigma_{1}\p_{0}X^i\Pi^i-\Lag$ can be written
by inverting (\ref{ConjMomentum}) :
\begin{equation}
 \Ham=\Ham_{Free} +\Ham_{2} + \Ham_{3}+O\left( \frac{1}{R^5} \right),
\end{equation}
with
\begin{eqnarray}
 \Ham_{Free} & = & \frac{1}{2}\int d\sigma_1 \left(\Pi^{2}+\left( \p_{1}X\right) ^{2}\right), \nonumber \\
\Ham_2 & = & \frac{c_2}{4} \int d\sigma_1 \left[\left(\Pi\cdot\Pi\right)^2+\left(\p_1X\cdot\p_1X\right)^2-2\left(\Pi\cdot\Pi\right)
\left(\p_1X\cdot\p_1X\right)\right], \\
\Ham_3 & = & \frac{c_3}{4} \int d\sigma_1 \left[\left(\Pi\cdot\Pi\right)^2+\left(\p_1X\cdot\p_1X\right)^2-2\left(\Pi\cdot\p_1X\right)^2\right].
\nonumber
\end{eqnarray}

To calculate the energy levels we consider $\Ham_{2,3}$ as a perturbation to the free Hamiltonian. We perform a Fourier expansion of the fields and their conjugate momenta at some given
time (say $\sigma_0=0$),
\begin{eqnarray}
 X^i(\sigma_1)&=&\frac{1}{\sqrt{\pi}}\sum_{n \neq 0}\frac{1}{n}\alpha_{n}^i
 \sin\left(\frac{\pi n\sigma_{1}}{R}\right),\\
\Pi^i(\sigma_1)&=&-\frac{i\sqrt{\pi}}{R}\sum_{n \neq 0}\alpha_{n}^i
\sin\left(\frac{\pi n\sigma_{1}}{R}\right). \nonumber
\end{eqnarray}
Note that while this expansion looks like an expansion of two independent operator-sets, $X(\sigma_1)$ and $\Pi(\sigma_1)$,  in terms of just one ($\alpha_n$),
the coefficients in the expansion of $X$ are actually $\alpha_n+\alpha_{-n}$, while
those in the expansion of $\Pi$ are $\alpha_n-\alpha_{-n}$, so that the two expansions
involve {\it independent} operators.
The canonical commutation relation $\left[X^i(\sigma_1) , \Pi^j(\sigma_1')\right]=i\delta^{ij}\delta\left( \sigma_1 -\sigma_1'\right)$ implies that the modes satisfy the commutation relation $\left[\alpha_{n}^{i},\alpha_{m}^{j}\right]=n\delta_{n,-m}\delta^{ij}$. Plugging this mode expansion into our Hamiltonians gives the
free (unregularized) Hamiltonian :
\begin{equation}
\label{eq:FreeHam}
 \Ham_{Free}=\frac{1}{2}\frac{\pi}{R}\sum_{n=-\infty}^{\infty}\alpha_{n}^i\alpha_{-n}^i,
\end{equation}
and the perturbations (keeping the $\sigma_1$ integration) :
\begin{equation}
\label{eq:DefHam}
 \Ham_{2,3} = \frac{c_{2,3}}{4} \frac{\pi^2}{R^4} \sum_{m,n,p,q \neq 0}\alpha_m^i\alpha_n^i\alpha_p^j\alpha_q^j
 \,
 \int_0^R d\sigma_1 \left[  S_n S_m S_p S_q +  C_n C_m C_p C_q  +2 H^{(2,3)}_{nmpq}\right],
\end{equation}
where $S_n  \equiv  \sin (n\pi \sigma_1 / R)$, $C_n  \equiv  \cos (n\pi \sigma_1 / R)$, and we define $H^{(2)}_{nmpq} =  S_n S_m C_p C_q$ and
 $H^{(3)}_{nmpq} = S_n C_m S_p C_q$.

The result of the $\sigma_1$ integral is rather lengthy. However, in this section we will
only be interested in the first order in perturbation theory in the deformations ${\cal H}_{2,3}$. At leading order in perturbation theory, only terms in the perturbation Hamiltonian that relate
degenerate energy levels of the free theory can contribute, which means that it is enough to
consider the terms in (\ref{eq:DefHam}) that have $n+m+p+q=0$ (note that $\alpha_{-n}$ increases the energy in the free theory by $n \pi / R$).
Considering only these zero-energy diagonal terms, which we denote $\Ham_*^d$, we have :
\begin{eqnarray}
\label{Hamiltonians}
 \Ham_{2}^d&= &\frac{c_{2}}{4}\frac{\pi^{2}}{R^{3}}\sum_{m,n\neq 0}\alpha_{m}^{i}\alpha_{n}^{i}\alpha_{-m}^{j}\alpha_{-n}^{j}, \\
\Ham_{3}^d&= &\frac{c_{3}}{4}\frac{\pi^{2}}{2R^{3}}\sum_{m,n \neq 0} \left( \alpha_{m}^{i}\alpha_{-m}^{i}\alpha_{n}^{j}\alpha_{-n}^{j}+
\alpha_{m}^{i}\alpha_{n}^{i}\alpha_{-m}^{j}\alpha_{-n}^{j} \right). \nonumber
\end{eqnarray}

Our discussion until now was classical and we ignored operator ordering issues, but quantizing the theory leaves us with an ordering ambiguity in $\Ham$.
The prescription we choose to deal with this is {\it Weyl ordering} -- averaging over all orderings, e.g.
$\alpha_n \alpha_m~\rightarrow~\left( \alpha_n\alpha_m +\alpha_m \alpha_n \right)/2!$, and similarly for four or more operators. Using this prescription, and then normal
ordering the resulting operators so that creation operators are to the right of
annihilation operators, yields~:
\begin{eqnarray}
\sum_{m,n\neq0}\alpha_{m}^{i}\alpha_{n}^{i}\alpha_{-m}^{j}\alpha_{-n}^{j}&\rightarrow&  (D-2) \sum_{n=1}^{\infty}n\sum_{m=1}^{\infty}m +
\frac{\left(D-2\right)^{2}}{2}\sum_{n=1}^{\infty}n^{2}+\frac{\left(D-2\right)}{2}\sum_{n=1}^\infty n^{2}+  \nonumber \\
&& \sum_{n=1}^\infty \left(4\sum_{m=1}^\infty m+2n\left(D-2\right)+2n\right)\alpha_{-n}^{i}\alpha_{n}^{i}+ \\
&& 2\sum_{m,n=1}^\infty \left( \alpha_{-m}^{i}\alpha_{-n}^{i}\alpha_{m}^{j}\alpha_{n}^{j}+\alpha_{-m}^{i}\alpha_{-n}^{j}\alpha_{m}^{i}\alpha_{n}^{j} \right),  \nonumber  \\
\sum_{m,n\neq0}\alpha_{m}^{i}\alpha_{-m}^{i}\alpha_{n}^{j}\alpha_{-n}^{j}&\rightarrow& (D-2)^2 \sum_{m,n=1}^{\infty}mn+(D-2) \sum_{n=1}^{\infty}n^{2}+
4\sum_{m,n=1}^{\infty}\alpha_{-m}^{i}\alpha_{-n}^{j}\alpha_{m}^{i}\alpha_{n}^{j}+
\nonumber \\ &&
\sum_{n=1}^\infty4\left(n+(D-2) \sum_{m=1}^\infty m\right)\alpha_{-n}^{i}\alpha_{n}^{i} . \nonumber
\end{eqnarray}
This contains several divergent sums, that we regularize with a zeta function regularization :
\begin{equation}
 \sum_{n=1}^\infty n = -\frac{1}{12}, \quad \sum_{n=1}^\infty n^2 = 0, \quad \sum_{n=1}^\infty n^3  = \frac{1}{120} , \quad \sum_{n=1}^\infty n^4 = 0.
\end{equation}
We then obtain :
\begin{eqnarray} \label{regsums}
\sum_{m,n\neq0}\alpha_{m}^{i}\alpha_{n}^{i}\alpha_{-m}^{j}\alpha_{-n}^{j}&\rightarrow&  \frac{D-2}{12^2}+
  \sum_{n=1}^\infty\left(-\frac{1}{3}+2n\left(D-1\right)\right)\alpha_{-n}^{i}\alpha_{n}^{i}+\\
  && 2\sum_{m,n=1}^\infty\left(\alpha_{-m}^{i}\alpha_{-n}^{i}\alpha_{m}^{j}\alpha_{n}^{j}+\alpha_{-m}^{i}\alpha_{-n}^{j}\alpha_{m}^{i}\alpha_{n}^{j}\right), \nonumber \\
\sum_{m,n\neq0}\alpha_{m}^{i}\alpha_{-m}^{i}\alpha_{n}^{j}\alpha_{-n}^{j}&\rightarrow& \frac{\left(D-2\right)^{2}}{12^2}+
4\sum_{m,n=1}^{\infty}\alpha_{-m}^{i}\alpha_{-n}^{j}\alpha_{m}^{i}\alpha_{n}^{j}+
 \sum_{n=1}^\infty4\left(n-\frac{D-2}{12 }  \right)\alpha_{-n}^{i}\alpha_{n}^{i} . \nonumber
\end{eqnarray}
The normal-ordered free Hamiltonian (\ref{eq:FreeHam}) now takes the form
\begin{equation}
\Ham_{Free}=\frac{\pi}{R}\left[\sum_{n=1}^{\infty}\alpha_{-n}^i\alpha_{n}^i- \frac{D-2}{24}\right].
\end{equation}
Plugging (\ref{regsums}) into the perturbation Hamiltonians (\ref{Hamiltonians}) gives us a regularized normal ordered form
from which we can calculate the energy levels.

As in \cite{LW}, we consider the $O\left( 1 / R^3\right) $ corrections to the lowest-lying energy levels of the free theory, which can be decomposed into representations of $SO(D-2)$. We annotate the low-lying states as follows~:
\begin{center}
\begin{tabular}{|llll|}
\hline
$\ket{n,i}$ & O(D-2) multiplet & Fock representation & Degeneracy $\omega_{n,i}$\\
\hline
$\ket{0}$ & Scalar & $\ket{0}$ & $1$\\
$\ket{1}$ & Vector & $\alpha_{-1}^k\ket{0}$ & $D-2$\\
$\ket{2,1}$ & Scalar & $ \alpha_{-1}^k \alpha_{-1}^k \ket{0} $ & $1$\\
$\ket{2,2}$ & Vector & $\alpha_{-2}^k\ket{0}$ & $D-2$\\
$\ket{2,3}$ & Symmetric 2-tensor & $\left( \alpha_{-1}^k \alpha_{-1}^l  -\frac{\delta^{kl}}{D-2} \alpha_{-1}^m \alpha_{-1}^m\right)\ket{0} $ &$\frac{1}{2}D(D-3)$\\
$\ket{3,2}$ & Vector & $\alpha_{-3}^k\ket{0}$ & $D-2$\\
\hline
                                                                                                              \end{tabular},
                                                                                                              \end{center}
where $n$ is the usual level related to the energy of the free theory (\ref{FreeEnergy}), and $i$ is an index going over the different states
at each level.

We denote by $E_{n,i}^0$ the energy levels up to $O\left( 1/R\right)$, arising from the free
Hamiltonian, and by $E_{n,i}^1$ the $O\left( 1/R^3\right)$ corrections arising from $c_{2,3}$. By computing the expectation value
$\left<n,i| \Ham_{Free}+\Ham_2+\Ham_3 |n,i\right>$ we obtain the leading energy levels with the L\"uscher term,
\begin{equation}
\label{FreeEnergy}
 E_n^0 = \frac{\pi}{R}\left[n-\frac{D-2}{24}\right],
\end{equation}
and the corrections
\begin{center}
\begin{tabular}{|ll|}
\hline
$\ket{n,i}$ & $E_{n,i}^1$\\
\hline
$\ket{0}$ & $\frac{\pi^{2}}{\left(24\right)^{2}R^{3}}\frac{\left(D-2\right)}{2}\left[2c_{2}+c_{3}\left(D-1\right)\right]$ \\
$\ket{1}$ & $E_{0}^1+\frac{\pi^{2}}{24R^{3}}\left[c_{2}\left(12D-14\right)+c_{3}\left(5D+7\right)\right]$ \\
$\ket{2,1}$ & $E_0^1+\frac{\pi^{2}}{R^{3}}\left[\left( 2c_2+c_3\right)\frac{11D+13}{12} -2c_2\left(2-\frac{D-2}{12} \right) \right]$\\
$\ket{2,2}$ & $E_0^1+\frac{\pi^{2}}{R^{3}}\left[\left( 2c_2+c_3\right)\frac{11D+13}{12} -2c_2\left(2-\frac{D-2}{12} \right) \right]$ \\
$\ket{2,3}$ & $E_0^1+\frac{\pi^{2}}{R^{3}}\left[\left( 2c_2+c_3\right)\frac{5D+25}{12} -2c_2\left(2-\frac{D-2}{12} \right) \right]$ \\
$\ket{3,2}$ & $E_0^1+\frac{\pi^{2}}{R^{3}}\left[ 3\left(\frac{D-2}{12}-3\right)c_{2}+\frac{1}{8}\left(17D+19\right)\left(c_{3}+2c_{2}\right)\right] $ \\
\hline
\end{tabular}
\end{center}
These exactly coincide with the results presented in \cite{LW} for these states,  confirming that our regularization scheme is equivalent to theirs.
As shown in \cite{LW,AK,AKS} the coefficients $c_{2,3}$ are actually constrained by the Lorentz symmetry $SO(1,D-1)$ of the underlying theory, which
imposes $c_2=\frac{1}{2T}$ , $c_3=-\frac{1}{T}$. This turns out to simplify $\Ham_2+\Ham_3$ considerably:
\begin{eqnarray}
 \Ham_2^d+\Ham_3^d& =& -\frac{\pi^2}{2 T R^3}\left[ \left( \sum_{n=1}^\infty \alpha_{-n}^i \alpha_n^i\right)^2
-\frac{D-2}{12}\sum_{n=1}^\infty \alpha_{-n}^i \alpha_n^i +\frac{\left(D-2 \right)^2}{24^2}\right] \\
 &=& -\frac{\pi^2}{2 T R^3}\left[ \sum_{n=1}^\infty \alpha_{-n}^i \alpha_n^i
-\frac{\left(D-2 \right)}{24}\right]^2, \nonumber
\end{eqnarray}
and one can see that this doesn't lift the degeneracy between states in the same level\footnote{It would be interesting to understand in the static gauge how the full Nambu-Goto Hamiltonian
gives rise to the expected energy levels (\ref{ngenergies}), and does not lift the degeneracies coming from
the free Hamiltonian. One may guess that this would happen by having the full Nambu-Goto
Hamiltonian be a function just of $\left( \sum\alpha_{-n}\cdot\alpha_{n}\right)$, but this
is not true already for $\Ham_2+\Ham_3$ when we look at the off-diagonal terms, so it is clear
that the eigenstates of the full Nambu-Goto Hamiltonian differ from those of the free theory.}.
The leading corrections to energy levels take the simple form
\begin{equation}
E_n^1 = -\frac{\pi^2}{2 T R^3}\left[ n
-\frac{\left(D-2 \right)}{24}\right]^2,
\end{equation}
in agreement with (6.1) in \cite{LW} and with the long string expansion of the Nambu-Goto open
string energy levels \cite{Arvis} (up to the constant $TR$ term, which we ignored in our work) : 
\begin{eqnarray}
\label{ngenergies}
E_n&=&\sqrt{ \left(TR\right)^2+2\pi T \left(n-\frac{D-2}{24} \right)} \\ &\approx& TR +\frac{\pi}{R}\left(n-\frac{D-2}{24} \right)-
\frac{\pi^2}{2TR^3}\left[n-\left(\frac{D-2}{24} \right) \right]^2+O\left( \frac{1}{R^5}\right).
\nonumber
\end{eqnarray}

\subsection{The corrections from the $c_4$ term}
\label{sec:OpenC4}

We now compute the leading corrections to the energy levels coming from $S_4$, which will give a $O\left( 1 / R^5\right)$ contribution to the energy levels. These are the leading corrections
to Nambu-Goto energy levels from bulk terms; we will discuss the leading corrections from
boundary terms in the next subsection. The $S_4$ term may be written in Minkowski space as
\begin{equation}
 S_4=-\int d^2\sigma \, c_{4}\left( \p_{0}^{2}X^i\p_{0}^{2}X^i-2\p_{0}\p_{1}X^i\p_{0}\p_{1}X^i+\p_{1}^{2}X^i\p_{1}^{2}X^i\right)
\left(\p_{1}X^j\p_1X^j-\p_{0}X^j\p_0X^j\right).
\end{equation}

Note that our action now includes terms with higher time derivatives, so we seem
to get new degrees of freedom at high energies, but these lie outside the low-energy approximation so we can ignore them.
However, we still need to get rid of terms with higher time derivatives in order to use the Hamiltonian formalism. We can do
this by adding to the action terms proportional to the equation-of-motion;
this does not modify the energy levels, since these terms can be canceled by a
field redefinition. Thus, we can use the (full) equation-of-motion to replace in our Hamiltonian
\begin{equation}
 \p_0^2 X^i \to \p_1^2 X^i + c_{2,3} \p^4 X^3 \cdots.
\end{equation}
 Up to the order in the derivative expansion that we work in, we can just keep the first term (the free equation-of-motion). This gives us the Hamiltonian (not forgetting
the correction to the conjugate momentum $\Pi^i$ from the $c_4$ term) :
\begin{equation}
 \Ham_4=-2c_{4}\int_0^R d\sigma_{1}\left(\p_{1}^{2}X\cdot \p_{1}^{2}X-\left(\p_{1}\Pi\right)^{2}\right)\left(\Pi^{2}-(\p_{1}X)^{2}\right).
\end{equation}

Repeating the procedure of the previous subsection, keeping again only terms that do not
change the energy which is all we need to first order in $c_4$, we get the normal-ordered Hamiltonian in terms of creation and annihilation operators~:
\begin{equation}
 \Ham_4^d = -\frac{c_{4}\pi^{4}}{R^{5}}\left\{ \sum_{m,n=1}^{\infty}mn\left[-4\alpha_{-m}^{i}\alpha_{-n}^{i}\alpha_{m}^{j}\alpha_{n}^{j}+4\alpha_{-m}^{i}\alpha_{-n}^{j}\alpha_{m}^{i}\alpha_{n}^{j}\right]+
\sum_{n=1}^{\infty} 4n^{3}\left(D-3\right)\alpha_{-n}^{i}\alpha_{n}^{i}\right\}.
\end{equation}
For the open string this gives us the following corrections to the energy levels ($E_{n,i}^2$ now denotes the $O\left( 1/R^5 \right)$ correction)~:
\begin{center}
\begin{tabular}{|ll|}
\hline
$\ket{n,i}$ & $E_{n,i}^2$\\
\hline
$\ket{0}$ & $0$ \\
$\ket{1}$ & $-4c_4\frac{\pi^4}{R^5}\left(D-3\right)$ \\
$\ket{2,1}$ & $0$\\
$\ket{2,2}$ & $-64 c_4\frac{\pi^4}{R^5}\left(D-3\right)$ \\
$\ket{2,3}$ & $-16c_{4}\frac{\pi^{4}}{R^{5}}\left(D-2\right)$ \\
\hline
\end{tabular}
\end{center}
Note that the degeneracy between different states at the same level is lifted, but there is
no correction to the ground state energy.

\subsection{Corrections to energy levels from boundary terms}
\label{sec:OpenB2}

As mentioned above, the leading allowed boundary term for Dirichlet boundary conditions is~:
\begin{equation}
\label{b2term}
 S'_2=b_2\int d^2\sigma \, \left[ \delta\left(\sigma_1 - R\right)+\delta\left(\sigma_1\right) \right]\left(\p_{0}\p_{1}X \cdot \p_{0}\p_{1}X\right).
\end{equation}

Repeating our procedure with this interaction term gives the Hamiltonian (keeping only zero
energy terms) :
\begin{equation}
 (\Ham_{2}')^d = -b_{2}\frac{\pi^{3}}{R^{4}}\left(4\sum_{n=1}^{\infty} n^{2}\alpha_{-n}^{i}\alpha_{n}^{i}+\frac{D-2}{60}\right),
\end{equation}
which gives the following contributions to the energy levels :
\begin{center}
\begin{tabular}{|ll|}
\hline
$\ket{n,i}$ & ${E'}_{n,i}^1$\\
\hline
$\ket{0}$ & $\frac{-b_{2}\pi^{3}}{R^{4}}\frac{D-2}{60}$ \\
$\ket{1}$ & ${E'}_0^1-4b_{2}\frac{\pi^{3}}{R^{4}}$ \\
$\ket{2,1}$ & ${E'}_0^1-8b_{2}\frac{\pi^{3}}{R^{4}}$\\
$\ket{2,2}$ & ${E'}_0^1-32b_{2}\frac{\pi^{3}}{R^{4}}$ \\
$\ket{2,3}$ & ${E'}_0^1-8b_{2}\frac{\pi^{3}}{R^{4}}$ \\
\hline
\end{tabular}
\end{center}
As mentioned above, for large $R$ these are the leading corrections to the Nambu-Goto energy levels if $b_2\neq0$. In this case there is a non--zero correction also to the ground
 state energy, and again we see that some of the degeneracy is lifted.


\subsection{Corrections to average energy}

In order to verify the consistency of our results, we can compare our results to the partition
function of the effective string theory on the annulus \cite{LW,AK,AF}. This partition
function, in the open string channel, does not include all the information about the energy
levels, but it includes the averaged energy at each level of the free theory, $\left[E_n\right] \equiv \sum_i \frac{\omega_{n,i}}{\omega_n}E_{n,i}$.

To compute the partition function we work in Euclidean space and compactify the $\sigma_0$ ($X^0$) direction with period $L$. 
The partition function on the annulus may be written in the open-string channel as~:
\begin{equation}
 Z^{open,annulus}=\sum_{n,i}\omega_{n,i} e^{-E_{n,i}\left(R\right)L}.
\end{equation}
By expanding in orders of $1/R$, we can identify the corrections arising from each term in the action. Considering first the bulk action, we have up to the order we work in :
\begin{eqnarray}
 Z^{open,annulus}&=&\sum_{n}\omega_{n} e^{-E_{n}^0\left(R\right)L}\left[ 1- L\left[ E_{n,i}^1\right]  + \frac{1}{2}L^2\left[\left( E_{n,i}^1 \right)^2\right]
-L\left[E_{n,i}^2\right]  \right] \\
&=& Z_0\left[1- \left< S_{2,3}\right> + \frac{1}{2}\left< \left( S_{2,3}\right)^2 \right> -\left<S_4\right>  \right]. \nonumber
\end{eqnarray}
 The $\left< S _i\right>$'s were given in \cite{LW,AK} in terms of Eisenstein series of $q=e^{-\pi\frac{L}{R}}$. By equating the different coefficients
of $q^n$ we can relate the average energy shifts to the $\left< S _i\right>$'s. We can extract the $c_4$ corrections by looking at $\langle S_4 \rangle$ in the partition function
approach~:
\begin{eqnarray}
\label{OpenCorrections}
 \left[E_0^{open,c_4}\right]&=& 0, \nonumber \\
 \left[E_1^{open,c_4}\right]&=& -4c_{4}\frac{\pi^{4}}{R^{5}}\left(D-3\right),\\
\left[E_2^{open,c_4}\right]&=& -4c_{4}\frac{\pi^{4}}{R^{5}}\frac{1}{\omega_{2}}\left(D-2\right)\left(D-3\right)\left(16+D\right). \nonumber
\end{eqnarray}
By summing over the state degeneracies (with $\omega_2 \equiv \sum_i \omega_{2,i}$)
we obtain from our results in section \ref{sec:OpenC4} exactly the same corrections as (\ref{OpenCorrections}).

The same can be done for the boundary term (\ref{b2term}), whose effect on the partition
 function was computed in \cite{AF}; the partition function gives :
\begin{eqnarray}
 \left[E_0^{open,b_2}\right]&=& -\frac{b_{2}\pi^{3}}{R^{4}}\frac{D-2}{60}, \nonumber \\
\left[E_1^{open,b_2}\right]&=& -4b_{2}\frac{\pi^{3}}{R^{4}} + E_0^{open,b_2},  \\
\left[E_2^{open,b_2}\right]&=& -4b_{2}\frac{\pi^{3}}{R^{4}\omega_2}\left(D-2\right)\left(D+7\right) + E_0^{open,b_2}, \nonumber
\end{eqnarray}
which are again identical to the weighted averaged corrections one calculates from our results in section \ref{sec:OpenB2}.

\section{Closed string energy levels}

We now turn to the closed string states, compactifying $\sigma^1$ ($X^1$) with period $r$.
The mode expansion now takes the form (ignoring the zero modes, since we can focus on
states with zero transverse momentum) :
\begin{eqnarray}
 X^i(\sigma_1) &=&\frac{i}{2\sqrt{\pi}}\sum_{n\neq0}\frac{1}{n}\left[\alpha_{n}^i e^{-i\frac{2\pi n \sigma_{1}}{r}}+\tilde{\alpha}_{n}^i
e^{i\frac{2\pi n \sigma_{1}}{r}}\right], \nonumber \\
\Pi^i(\sigma_1) &=&\frac{\sqrt{\pi}}{r}\sum_{n \neq 0}\left[\alpha_{n}^ie^{-i\frac{2\pi n \sigma_{1}}{r}}+\tilde{\alpha}_{n}^ie^{i\frac{2\pi n \sigma_{1}}{r}}\right].
\end{eqnarray}
Note that $X^i$ is now expanded in the modes $\alpha_n^i-\tilde\alpha_{-n}^i$, and
$\Pi^i$ is expanded in $\alpha_n^i+\tilde\alpha_{-n}^i$, so that the two operators have
independent expansions as before.
The commutation relation of $X$ and $\Pi$ now implies (if we assume also a worldsheet parity
symmetry exchanging $\alpha_n^i$ with $\tilde\alpha_n^i$)
$\left[\alpha_n^i,\alpha_m^j \right]=\left[\tilde{\alpha}_n^i,\tilde{\alpha}_m^j \right]=n\delta_{n,-m}\delta^{ij}$,
$\left[{\alpha}_n^i,\tilde{\alpha}_m^j \right]=0$.

In terms of this mode expansion, using the same ordering prescription and regularization as in the previous section, and using the physical values of $c_2$ and $c_3$, the $c_{2,3}$ Hamiltonians give us the following correction to the free Hamiltonian
at leading ($1/r^3$) order (keeping again just the zero-energy terms) :
\begin{equation} \label{htwothree}
 \Ham_{2,3}^d= -\frac{\pi^2}{2Tr^3}\left[\frac{\left(D-2\right)^{2}}{6^{2}}-
\frac{4\left(D-2\right)}{6}\sum_{n=1}^{\infty} \left(\alpha_{-n}^i \alpha_n ^i + \tilde{\alpha}_{-n}^i \tilde{\alpha}_n ^i \right)+
16 \sum_{n,m=1}^{\infty} \alpha_{-n}^i \alpha_n ^i \tilde{\alpha}_{-m}^j \tilde{\alpha}_m ^j\right].
\end{equation}
This agrees with the leading correction to the Nambu-Goto energy levels; if we denote by
$N_L$ the level of the left-moving oscillators $\tilde\alpha_n$, and by $N_R$ the level of the right-moving oscillators $\alpha_n$, the full Nambu-Goto answer takes the form
\begin{equation}
E_{N_L,N_R}(r)^2 = (T r)^2 + 4 \pi T \left(N_L + N_R - \frac{D-2}{12}\right) + \left( \frac{2\pi (N_L-N_R)}{r} \right)^2,
\end{equation}
and expanding this in powers of $1/r$ reproduces the result from (\ref{htwothree}).

The zero-energy terms in the first correction to Nambu-Goto for closed strings $\Ham_4$ now take the form
\begin{equation}
 \Ham_{4}^d= \frac{128\pi^{4}}{r^{5}}c_{4}\sum_{m,n\neq0}mn\alpha_{n}^{i}\alpha_{-n}^{j}
 \tilde{\alpha}_{m}^{i}\tilde{\alpha}_{-m}^{j}.
\end{equation}
Repeating our ordering and regularizing procedure gives
\begin{equation}
 \Ham_4^d= \frac{128\pi^{4}}{r^{5}}c_{4}\sum_{m,n=1}^{\infty}mn\left(\alpha_{-n}^{j}\alpha_{n}^{i}-\alpha_{-n}^{i}\alpha_{n}^{j}\right)
\left(\tilde{\alpha}_{-m}^{j}\tilde{\alpha}_{m}^{i}-\tilde{\alpha}_{-m}^{i}\tilde{\alpha}_{m}^{j}\right).
\end{equation}
This Hamiltonian looks like a spin-spin interaction between the left and right movers, and it's clear that pure-left or pure-right states
will get no energy contribution; therefore we treat only the mixed states in this calculation. We also note that in $D=3$, $\Ham_4=0$ identically
(as expected, since the $c_4$ term is a total derivative in this case).
We annotate the lowest states containing both right- and left-movers as follows :
\begin{center}
\begin{tabular}{|lll|}
\hline
$\ket{n,i}$ & State & Degeneracy $\omega_{n,i}$\\
\hline
$\ket{2,1}$ & $\alpha_{-1}^{k}\tilde{\alpha}_{-1}^{k}\ket{0}$ & $1$\\
$\ket{2,2}$ & $\left(\alpha_{-1}^{k}\tilde{\alpha}_{-1}^{l}+\alpha_{-1}^{l}\tilde{\alpha}_{-1}^{k}-\frac{2\delta^{kl}}{D-2}\alpha_{-1}^{j}\tilde{\alpha}_{-1}^{j}\right)\ket{0}$ &
 $\frac{1}{2}D(D-3)$\\
$\ket{2,3}$ & $\left(\alpha_{-1}^{k}\tilde{\alpha}_{-1}^{l}-\alpha_{-1}^{l}\tilde{\alpha}_{-1}^{k}\right)\ket{0}$ &
 $\frac{1}{2}(D-2)(D-3)$\\
\hline
\end{tabular}
\end{center}
Then, the leading order contribution of $\Ham_4$ to the energy levels is
\begin{center}
\label{ClosedString}
\begin{tabular}{|ll|}
\hline
$\ket{n,i}$ & $E_{n,i}^2$\\
\hline
$\ket{2,1}$ & $\mathcal{E}\left(D-3\right)$\\
$\ket{2,2}$ & $-\mathcal{E}$ \\
$\ket{2,3}$ & $\mathcal{E}$ \\
\hline
\end{tabular},
\end{center}
with $\mathcal{E}\equiv 256 \pi^{4} c_4 / r^{5}$, and with no correction to lower states or
to other states at level $2$.

\subsection{Corrections to average energy}

We can now compare our results to both partition functions considered in \cite{AK} -- the torus and the annulus.
For the torus it was shown that there is no
contribution to the closed string {\it averaged} energy at leading order in $c_4$ (for any $D$). We can check this by calculating from the results above
\begin{equation}
 \left[E_2^{closed,c_4}\right] = \sum_i \frac{\omega_{2,i}}{\omega_2}E_{2,i}^2 =0.
\end{equation}

For the annulus, among the level 2 states, only the singlet state $\ket{2,1}=\alpha_{-1}^k\tilde{\alpha_{-1}^k}\ket{0}$ has an overlap with the boundary, so
we expect $\left[ E_2 ^{annulus, c_4}\right]= E_{2,1}^2$. Indeed, \cite{AK} showed (by expanding the partition function in powers of
$\tilde{q}^{ann.}=e^{-\frac{4\pi L}{r}}$) that
\begin{equation}
 \left[ E_2 ^{annulus, c_4}\right]=\frac{256 \pi^{4}}{r^{5}}c_{4}\left(D-3\right),
\end{equation}
which is in agreement with $E_{2,1}^2$ found above.

\subsection{Higher-order contributions to the closed-string ground-state energy}

As we saw, the $c_4$ term doesn't contribute at linear order to the ground-state energies, nor to
any states for $D=3$, so it is interesting to look for the leading corrections to
these states.
In the closed string case, due to the commutativity of the left and right operators, we can easily calculate the contributions to
the ground-state energy due to higher derivative terms. Introducing light-cone coordinates for convenience, defined by
$\sigma_{\pm}=\left( \sigma_0 \pm \sigma_1 \right)/\sqrt{2}$, we consider two possible terms that are not part of the derivative-expansion of
the Nambu-Goto action, and that can appear at 8-derivative order $O\left( 1/r^7\right)$ :
\begin{eqnarray}
\label{higherterms}
 S_9 \equiv c_{9}\int\p_{+}^{2}X^{i}\p_{+}^{2}X^{i}\p_{-}^{2}X^{j}\p_{-}^{2}X^{j}d^2\sigma ,\\
S_{10} \equiv c_{10}\int \p_{+}^{2}X^{i}\p_{-}^{2}X^{i}\p_{+}^{2}X^{j}\p_{-}^{2}X^{j}d^2\sigma . \nonumber
\end{eqnarray}

In the spirit of \cite{AK} we can show that these two terms are the only possible $\p^8 X^4$ terms one can write, up to terms which
contain the equation-of-motion. Note that the two terms coincide (but are non-zero) in $D=3$.
One may think that there are also $\p^8 X^6$ and $\p^8 X^8$ terms which contribute at the same
order $O\left( 1/r^7\right)$, but in fact using Lorentz invariance one can show that the former terms do not exist for $D=3$, and are related to the $c_4$
term for $D>3$, while the latter terms  must be equal to their values in the Nambu-Goto action
\cite{AKS}. In fact, Lorentz invariance of the
$c_4$ term imposes that up to $8$-derivative order the $\p^6 X^4$ and $\p^8 X^6$
terms are
\begin{equation}
\label{higherc4}
 S_4=4c_4\int d^2\sigma\p_+^2 X^i \p_-^2 X^i \left[\p_+ X^j \p_- X^j +\frac{1}{T}\left( \p_+X^j \p_-X^j\right)^2+
\frac{1}{T}\p_+X^j \p_+X^j\p_-X^k \p_-X^k\right].
\end{equation}
Inspecting the new terms (\ref{higherc4}) shows that they do not contribute to the ground-state energy, as their Hamiltonian contains $3$ left-movers and
$3$ right-movers.

Repeating our normal ordering procedure we get contributions to the ground-state energy :
\begin{eqnarray}
\label{rseven}
 E_0^{closed,c_9} &=&c_{9}\frac{4}{15^{2}}\frac{\pi^{6}}{r^{7}}\left(D-2\right)^{2}, \\
E_0^{closed,c_{10}} &=& c_{10}\frac{4}{15^{2}}\frac{\pi^{6}}{r^{7}}\left(D-2\right). \nonumber
\end{eqnarray}
For $D=3$ these are the only contributions at $O(1/r^7)$. For $D>3$ there could have been additional
contributions at this order proportional to $c_4 c_{2,3}$, but it seems that they cancel out between second order perturbation theory and first order perturbation theory using terms in the Hamiltonian whose coefficients are proportional to $c_4 c_{2,3}$, so that (\ref{rseven})
gives the leading corrections to the ground state energy in all dimensions.
The same terms (\ref{higherterms}) should also give the leading correction to the Nambu-Goto energy levels for
excited states for $D=3$.

Note that at least the $c_{10}$ term is expected to be arbitrary (unconstrained by Lorentz symmetry), since
it could come from a $\int \sqrt{h}R^2$ term -- $h_{\alpha \beta}$ being the induced metric and $R$ its scalar curvature -- in the un-gauge-fixed
effective string action, and this generically contains all terms permitted by symmetry.

\section{Comparison to lattice results}

Our results above apply to any effective string theory; of course the coefficients
that are not determined by Lorentz invariance are expected to be different in different
theories. It would be nice to compare the results above to analytic results for solitonic
strings (where the effective action can be computed perturbatively), but as far as we know
the relevant computations have not yet been performed. Alternatively, we can try to match our
results to numerical simulations of strings, which have been performed both in the context
of confining gauge theories in $2+1$ and $3+1$ dimensions, and for domain walls in $2+1$
dimensional theories. Unfortunately, almost all of these simulations are not precise enough
to see any deviations from the Nambu-Goto energy levels; moreover, we expect the radius
of convergence of the $1/r$ expansion that we are working in to be of order $r_c\sqrt{T}=\sqrt{4\pi\left(N_L+N_R-1/6 \right)}$ \footnote{This is estimated from
the Nambu-Goto expression for the closed string energy levels, using the radius of convergence $\lambda_c = 1/x$
of the Taylor series of $\sqrt{1+\lambda x}$ around $\lambda=0$.}, and almost all simulations
are performed for smaller values of $r$ where a comparison to our results is not expected
to be meaningful \footnote{One exception is \cite{Lattice1,Lattice2,Lattice3,Lattice4,Lattice5}, where a deviation from the Nambu-Goto
energy levels is reported, but this deviation seems to be at a lower order in $1/r$ than the one
predicted by Lorentz invariance; the source of this discrepancy is not clear.}.

Recently, a deviation from the Nambu-Goto energy levels was reported for some specific
closed string states in lattice simulations of $3+1$ dimensional $SU(N)$ Yang-Mills theory
\cite{Barak,Teper}, which calculated the spectrum for the low levels
of a closed flux-tube wrapping a compact dimension\footnote{See \cite{Teperreview} for a
review with references of previous work on this topic.}. Almost all the results there are still
for smaller radii than the expected radius of convergence, but nevertheless we can try to see if
the results there are consistent with the expected deviations. The lattice results are most
precise for the lightest state with given quantum numbers.

As discussed above, for closed strings in $D=4$ the leading shifts from Nambu-Goto for excited states are expected to come from $\Ham_4$. In order to compare these to the simulations, let
us discuss in more detail the quantum numbers that can be identified on the lattice for $D=4$.
$N_L$ and $N_R$ cannot be measured directly on the lattice, but their difference, $q = N_L-N_R$
which is the longitudinal momentum, can be measured.
The lattice breaks the $SO(2)$ transverse rotation group to $Z_4$, but the angular momentum
in this plane modulo $4$ can still be identified. Thus, it is useful to use a basis of states
which diagonalizes the transverse spin $J=J^{23}$,
\begin{equation}
\alpha_{n}^{\pm}\equiv\frac{\alpha^{2}\pm i\alpha^{3}}{\sqrt{2}}, \ \ \ \
\tilde{\alpha}_{n}^{\pm}\equiv\frac{\tilde{\alpha}^{2}\pm i\tilde{\alpha^{3}}}{\sqrt{2}},
\end{equation}
such that the spin of these states is
$J^{23}\alpha^{\pm}_{-n}\ket{0}=\pm\alpha^{\pm}_{-n}\ket{0}$, $J^{23}\tilde{\alpha}^{\pm}_{-n}\ket{0}=\pm\tilde\alpha^{\pm}_{-n}\ket{0}$. The new operators
obey the commutation relations $\left[ \alpha, \tilde{\alpha}\right]=0$, $\left[ \alpha_n^{\pm},\alpha_m^{\pm}\right]=0$, and
$\left[ \alpha_n^{+},\alpha_m^{-}\right]=n\delta_{n,-m}$, and the same for the left-movers.
In terms of these, $\Ham_4$ takes the form
\begin{equation}
 \Ham_4=-\mathcal{E}\sum_{m,n=1}^{\infty}mn\left(\alpha_{-n}^{-}\alpha_{n}^{+}-\alpha_{-n}^{+}\alpha_{n}^{-}\right)
\left(\tilde{\alpha}_{-m}^{-}\tilde{\alpha}_{m}^{+}-\tilde{\alpha}_{-m}^{-}\tilde{\alpha}_{m}^{+}\right),
\end{equation}
and the lowest states whose energies get a correction to their Nambu-Goto value are (denoting $n=N_L+N_R$, $J=J^{23}$)~:
\begin{center}
\begin{tabular}{|lll|}
\hline
$\ket{n,q,J}$ & State & $E^2_{n,q,J}$\\
\hline
$\ket{2,0,0}$ & $\tilde{\alpha}_{-1}^+\alpha_{-1}^-\ket{0}$ & $\mathcal{E}$\\
$\ket{2,0,2}$ & $\tilde{\alpha}_{-1}^+\alpha_{-1}^+ \ket{0}$ & $-\mathcal{E}$\\
$\ket{3,1,0}$ & $\tilde{\alpha}_{-2}^+\alpha_{-1}^-\ket{0}$ & $4\mathcal{E}$\\
$\ket{3,1,1}$ & $\tilde{\alpha}_{-1}^+\tilde{\alpha}_{-1}^+\alpha_{-1}^-\ket{0}$ & $2\mathcal{E}$\\
$\ket{3,1,1}'$ & $\tilde{\alpha}_{-1}^+\tilde{\alpha}_{-1}^-\alpha_{-1}^+\ket{0}$ & $0$\\
$\ket{3,1,2}$ & $\tilde{\alpha}_{-2}^+\alpha_{-1}^+\ket{0}$ & $-4\mathcal{E}$\\
$\ket{3,1,3}$ & $\tilde{\alpha}_{-1}^+\tilde{\alpha}_{-1}^+\alpha_{-1}^+\ket{0}$ & $-2\mathcal{E}$\\

\hline
                                                                                                              \end{tabular}.
													      \end{center}
Here, as above, $\mathcal{E} = 256 \pi^4 c_4 / r^5$. The most prominent feature of these corrections is the
{\it splitting between different states in each level}. Note that we omitted some states
which are related by parity transformations (to be described below) to the states we wrote,
as they get the same correction.

\begin{figure}[ht!]
 \subfigure[]{
  {\begin{overpic}[width=0.55\linewidth,height= 12cm, trim=10 30 32 30, clip]{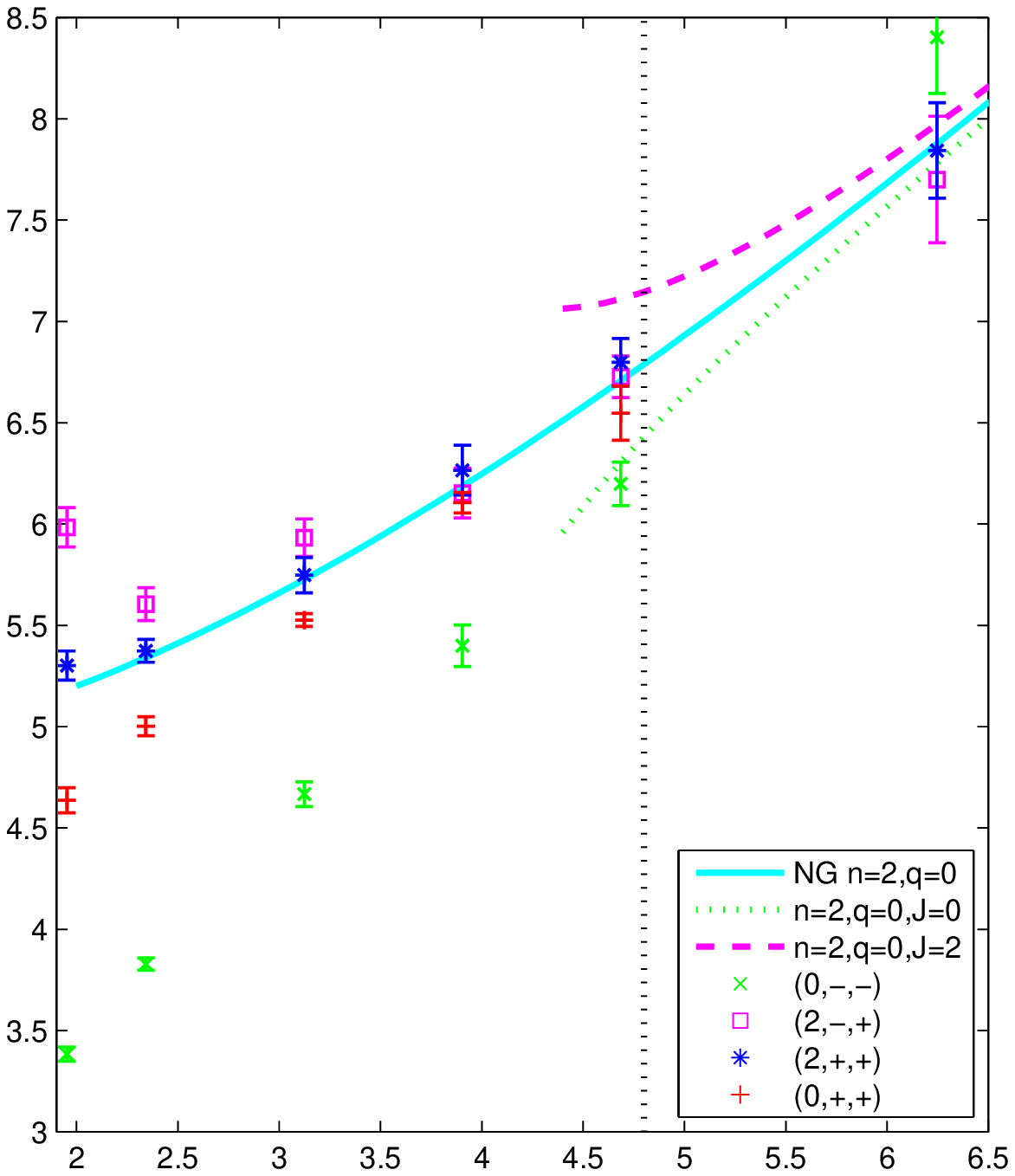}
  \put(27,1){$r\sqrt{T}$}
\put(-5,50){$E/\sqrt{T}$}
  \end{overpic}}
}
\subfigure[]{

  {\begin{overpic}[width=0.55\linewidth,height= 12cm, trim=10 35 35 35, clip]{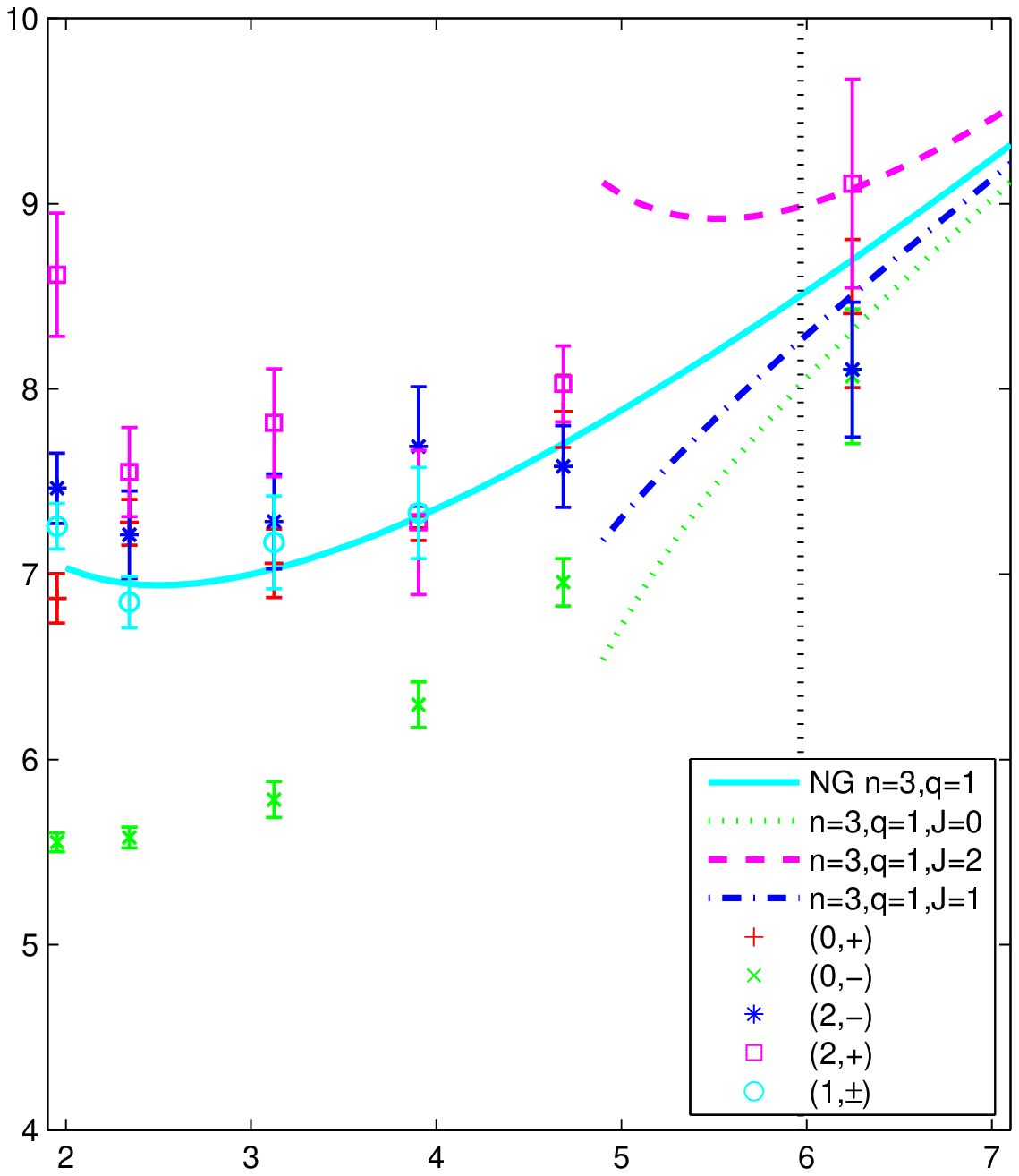}
  \put(30,1){$r\sqrt{T}$}
  \end{overpic}}

}

\caption{\textit {The expected energy levels, and the ones computed in lattice simulations of $SU(3)$ gauge theory \cite{Teper}, for level 2 with $q=0$ ($N_L=N_R=1$), and level 3 with $q=1$ ($N_L=2$, $N_R=1$). The discrete points are the lattice results from \cite{Teper},
annotated as $(|J|,P_\perp,P_\parallel)$ for $q=0$ and as  $(|J|,P_\perp)$ for $q=1$; the solid lines are the corresponding
Nambu-Goto energy levels, and the other lines
 include the shifts we calculated from $\Ham_4$ (for the lightest $n=2,3$ state with the given quantum
 numbers), using the specific value $c_4=(D-26)/192\pi^2 T^2$. The vertical line is the expected radius of convergence for each level, we expect a matching only for points
 that are well to the right of this line. }}
 \label{fig:NG}
\end{figure}

Two useful parity operators used in \cite{Barak,Teper} to classify the states are $P_\perp$ and $P_\parallel$. $P_{\perp}$ is a reflection about
the $X^2$ axis, which takes $\alpha^2 \rightarrow -\alpha^2$ or $\alpha^\pm \rightarrow -\alpha^{\mp}$, and $P_{\parallel}$ is the parity operator which
takes left $\rightarrow$ right. Using $P_\perp$ one can take $J$ states into $(-J)$ states, and since $\left[P_\perp,\Ham_4 \right]=0$ the negative $J$
states will be degenerate with the ones we wrote. Thus, in the continuum theory, $P_\perp$
 is only a useful quantum number for states with $J=0$, and our leading order correction
 does not distinguish between $J~=~0$ states with different values of $P_\perp$. Note, however,
 that in the lattice computation
$J$ is only well-defined modulo 4, so states with $J=\pm 2$ can also form combinations with
well-defined eigenvalues for $P_{\perp}$ (however, these must be degenerate in the continuum
limit). Similarly, $P_\parallel$ is a worldsheet parity transformation, exchanging $\alpha$ with $\tilde\alpha$ and taking $q\rightarrow -q$ (it also obeys $\left[P_\parallel,\Ham_4 \right]=0$).
Thus, it is only a useful quantum number for $q=0$ states.

Note that for $q=0$ the only state which is lower than the ones presented in the table above is the $N_L=N_R=0$ ground state with $J=0$,
 $P_\perp=P_\parallel=+$, so three of the four states with $n=2$ are
the lightest states in their sector. Similarly, for $q=1$ the only lighter states are the
$N_L+N_R=1$ states with $J=1$,
so the states with $J=0,2$ are the lightest states in their sector and can be measured
precisely on the lattice.

The results of \cite{Barak,Teper} show that even for rather small values of $r$ (where the
$1/r$ expansion is not expected to converge) almost all states agree very well with
the Nambu-Goto formula for the energy levels, except for two states (one with $q=0$
and one with $q=1$) that are the lightest states with $J^{P_{\perp}}=0^-$.
These states show a large deviation from Nambu-Goto for small values of $r$ where the
$1/r$ expansion does not converge; it is difficult to measure them at larger values of
$r$, and so far measurements for values of $r$ that are above the expected radius of
convergence of the $1/r$ expansion show no meaningful deviation from the Nambu-Goto
value (but they are also consistent with the expected deviations based on our results).

At the level of the leading deviation computed above, there is no
difference between the $0^-$ states and the other states, because as we saw the splitting is only $J$-dependent.
There are thus two possible interpretations of the results of \cite{Barak,Teper}. One possibility is that
when we sum the $1/r$ expansion (including all corrections coming from the $c_4$ term
and all terms related to it by Lorentz symmetry, and perhaps including additional
corrections as well) we will find a very small deviation for
almost all states, but a large deviation for the $0^-$ states. Another possibility discussed
in \cite{Teper} is that the deviation of the $0^-$ states from Nambu-Goto is related to a massive
particle with $0^-$ quantum numbers on the worldsheet; for large values of $r$ we would
expect such a massive particle to mix strongly with the massless modes and to decay almost
immediately, but for small enough values of $r$ this particle is lighter than the lightest
state with the same quantum numbers made from the massless modes, so there could be a state
which is dominantly made from this massive particle, and thus exhibits a large mismatch from
the Nambu-Goto expectations.

The results so far are not precise enough to distinguish between these possibilities; it would
be nice to have more accurate lattice results, and to be able to sum the $1/r$ expansion (at
least for some of the corrections) to increase its range of validity. Nevertheless, to illustrate
the issue we show in figure \ref{fig:NG} both the aforementioned lattice results, and the expected
energy levels computed in our $1/r$ expansion (which are meaningful only far to the right of the
vertical line in the figure, which is the estimated radius of convergence of the $1/r$ expansion). For the theoretical results we chose the value
$c_4= \left( D-26\right)/192 \pi T^2$, because the results
of \cite{AKS} imply that it is the universal value for consistent effective string actions,
even though it is not yet known how to see this directly in the static gauge.
From the figure it is clear that the present data can be matched smoothly to the
large $r$ theoretical results, but it is certainly also possible that it comes from
a massive state which is not captured by the effective string theory.

\begin{center}
\textbf{Acknowledgements}
\end{center}

We would like to thank M. Field, Z. Komargodski and A. Schwimmer for many interesting
discussions and for collaborations on related topics, and to thank B. Bringoltz, M. Caselle, A. Rajantie, J. Sonnenschein, and
M. Teper for useful discussions. OA would like to thank ECT*, Trento for hospitality during the conclusion of this work, and the participants in the ``Confining flux tubes and strings'' workshop there for useful discussions; this was supported in
part by the European Community Research Infrastructure Action under the
FP7 ``Capacities'' Specific Programme, project ``HadronPhysics2''.
This work was supported in part by the Israel--U.S.~Binational Science Foundation, by a research center supported by the Israel Science Foundation (grant number 1468/06), by a grant (DIP H52) of the German Israel Project Cooperation, and by the Minerva foundation with funding from the Federal German Ministry for Education and Research.

\bibliography{EnergyLevelsPaper}

\providecommand{\href}[2]{#2}\begingroup\raggedright\begin{thebibliography}{10}

\bibitem{LW}
M.~Luscher and P.~Weisz, ``{String excitation energies in SU(N) gauge theories
  beyond the free-string approximation},'' {\em JHEP} {\bfseries 07} (2004)
  014,
\href{http://arxiv.org/abs/hep-th/0406205}{{\ttfamily arXiv:hep-th/0406205}}.

\bibitem{AK}
O.~Aharony and E.~Karzbrun, ``{On the effective action of confining strings},''
  \href{http://dx.doi.org/10.1088/1126-6708/2009/06/012}{{\em JHEP} {\bfseries
  06} (2009) 012},
\href{http://arxiv.org/abs/0903.1927}{{\ttfamily arXiv:0903.1927 [hep-th]}}.

\bibitem{AKS}
O.~Aharony, Z.~Komargodski, and A.~Schwimmer. Work in progress, presented by O.
  Aharony at the Strings 2009 conference, June 2009, {\tt
  http://strings2009.roma2.infn.it/talks/Aharony\_Strings09.ppt}, and at the
  ECT* workshop on ``Confining flux tubes and strings'', July 2010, {\tt
  http://www.ect.it/Meetings/ConfsWksAndCollMeetings/ConfWksDocument/
  2010/talks/Workshop\_05\_07\_2010/Aharony.ppt}.

\bibitem{AF}
O.~Aharony and M.~Field, ``{On the effective theory of long open strings},''
\href{http://arxiv.org/abs/1008.2636}{{\ttfamily arXiv:1008.2636 [hep-th]}}.

\bibitem{Barak}
A.~Athenodorou, B.~Bringoltz, and M.~Teper, ``{The spectrum of closed loops of
  fundamental flux in D = 3+1 SU(N) gauge theories},''
\href{http://arxiv.org/abs/0912.3238}{{\ttfamily arXiv:0912.3238 [hep-lat]}}.

\bibitem{Teper}
A.~Athenodorou, B.~Bringoltz, and M.~Teper, ``{Closed flux tubes and their
  string description in D=3+1 SU(N) gauge theories},''
\href{http://arxiv.org/abs/1007.4720}{{\ttfamily arXiv:1007.4720 [hep-lat]}}.

\bibitem{Arvis}
J.~F. Arvis, ``{The exact q anti-q potential in Nambu string-theory},''
\href{http://dx.doi.org/10.1016/0370-2693(83)91640-4}{{\em Phys. Lett.}
  {\bfseries B127} (1983) 106}.

\bibitem{Lattice1}
M.~Caselle, M.~Hasenbusch, and M.~Panero, ``{Comparing the Nambu-Goto string
  with LGT results},''
  \href{http://dx.doi.org/10.1088/1126-6708/2005/03/026}{{\em JHEP} {\bfseries
  0503} (2005) 026}, \href{http://arxiv.org/abs/hep-lat/0501027}{{\ttfamily
  arXiv:hep-lat/0501027 [hep-lat]}}.

\bibitem{Lattice2}
M.~Billo and M.~Caselle, ``{Polyakov loop correlators from D0-brane
  interactions in bosonic string theory},'' {\em JHEP} {\bfseries 0507} (2005)
  038, \href{http://arxiv.org/abs/hep-th/0505201}{{\ttfamily
  arXiv:hep-th/0505201 [hep-th]}}.

\bibitem{Lattice3}
M.~Caselle, M.~Hasenbusch, and M.~Panero, ``{High precision Monte Carlo
  simulations of interfaces in the three-dimensional ising model: A comparison
  with the Nambu-Goto effective string model},'' {\em JHEP} {\bfseries 0603}
  (2006) 084, \href{http://arxiv.org/abs/hep-lat/0601023}{{\ttfamily
  arXiv:hep-lat/0601023 [hep-lat]}}.

\bibitem{Lattice4}
M.~Billo, M.~Caselle, and L.~Ferro, ``{The partition function of interfaces
  from the Nambu-Goto effective string theory},'' {\em JHEP} {\bfseries 0602}
  (2006) 070, \href{http://arxiv.org/abs/hep-th/0601191}{{\ttfamily
  arXiv:hep-th/0601191 [hep-th]}}.

\bibitem{Lattice5}
P.~Giudice, F.~Gliozzi, and S.~Lottini, ``{The confining string beyond the
  free-string approximation in the gauge dual of percolation},''
  \href{http://dx.doi.org/10.1088/1126-6708/2009/03/104}{{\em JHEP} {\bfseries
  0903} (2009) 104}, \href{http://arxiv.org/abs/0901.0748}{{\ttfamily
  arXiv:0901.0748 [hep-lat]}}.

\bibitem{Teperreview}
M.~Teper, ``{Large N and confining flux tubes as strings - a view from the
  lattice},''
\href{http://arxiv.org/abs/0912.3339}{{\ttfamily arXiv:0912.3339 [hep-lat]}}.

\end{thebibliography}\endgroup

\end{document}